\documentclass[11pt,a4paper]{article}
\usepackage[T1]{fontenc}
\usepackage[utf8]{inputenc}
\usepackage{authblk}
\usepackage{fullpage}
\usepackage{graphicx}
\usepackage{xcolor}
\usepackage{algorithm}
\usepackage{algpseudocode}
\usepackage{amssymb}
\usepackage{amsthm}
\usepackage{amsmath}
\usepackage{url}
\usepackage{bbold}
\usepackage{txfonts}
\usepackage[square,numbers,comma,sort&compress]{natbib} 
\usepackage[colorlinks,urlcolor=red,citecolor=red,linkcolor=blue]{hyperref}
\title{LaBonte's method revisited: An effective steepest descent method for micromagnetic energy minimization}
\author[1]{Lukas Exl\thanks{Corresponding author, \texttt{lukas.exl@fhstp.ac.at}}}
\author[1]{Simon Bance}
\author[1]{Franz Reichel}
\author[1]{Thomas Schrefl}
\author[2]{Hans Peter Stimming}
\author[2]{Norbert J. Mauser}

\affil[1]{University of Applied Sciences, Department of Technology, A-3100 St.Poelten, Austria}
\affil[2]{Wolfgang Pauli Institute c/o Fakult\"at f. Mathematik, Universit\"at Wien, A-1090 Vienna, Austria}

\begin{document}
\maketitle
\begin{abstract}
We present a steepest descent energy minimization scheme for micromagnetics. The method searches on a curve that lies on the sphere which keeps the magnitude of the magnetization vector constant. The step size is selected according to a modified   Barzilai-Borwein method. Standard linear tetrahedral finite elements are used for space discretization. For the computation of static hysteresis loops the steepest descent minimizer is faster than a Landau-Lifshitz micromagnetic solver by more than a factor of two. The speed up on a graphic processor is 4.8 as compared to the fastest single-core CPU implementation. 
\end{abstract}
\begin{center}
\end{center}

\section{\label{sec:intro}Introduction}
\subsection{Hysteresis in micromagnetics}
Energy application and the quest for rare-earth free or rare-earth reduced permanent magnets \cite{Gutfleisch2011,Skomski2013} renewed the interest in micormagnetics of permanent magnets
\cite{Sepehri-AminH.T.OhkuboS.NagashimaM.YanoT.ShojiA.KatoT.Schrefl2013}. Most state-of-the-art micromagnetic solvers integrate the Landau-Lifshitz Gilbert (LLG) equation of motion in time. However, the accessible time scale of micromagnetics simulations is the range of nanoseconds. This time scale is irrelevant for permanent magnet
applications. The measurement time for hysteresis loops of permanent magnets is in the range of seconds.
Therefore micromagnetic solvers that minimize the energy directly instead of solving a time dependent equation might be more suitable for the analysis for magnetization reversal in permanent magnets. 

Hysteresis in non-linear system results from the path formed by subsequent local minima. Kinderlehrer and Ma \cite{Kinderlehrer1994a} computed hysteresis loops in ferromagnets from the continuation of solutions for decreasing and increasing applied fields. Due to the  constraint that the magnetization vector keeps its length requires the solution of   a constrained non-linear optimization problem at each point of the hysteresis loop. 

\subsection{Minimization techniques}

Originally, LaBonte \cite{LaBonte1969a} computed equilibrium magnetic states in micromagnetics numerically. He used a finite difference scheme, in order to descretize the Gibbs free energy of a Bloch-wall in a ferromagnetic film. At each step of an iterative procedure he selected a computational cell and  rotated the magnetization vector of the cell in the direction of the effective field: The normalized magnetization vector is replaced by the normalized effective field. Kosavisutte and Hayashi \cite{Kosavisutte1996} showed that the original LaBonte method can be accelerated by the use of over- and under-relaxation. These methods rotate the magnetization towards the effective field. The effective field is proportional to the gradient of the energy. Therefore these methods are   steepest descent minimization schemes with  a particular choice of step-length and a subsequent normalization step. 

In unconstrained minimization the use of conjugate directions rather than steepest descent direction method improves the convergence. Cohen and co-workers \cite{Cohen1989} introduced a  conjugate gradient method for the computation of  molecular orientation of liquid crystals. The orientation vectors in liquid crystals have a fixed lengths similar to the magnetization in ferromagnets. There are two modifications to the classical conjugate gradient method: (1) The gradients are projected onto a plane normal to the current orientation vector. (2) After the line search the solution vector is normalized. In micromagnetics a similarly modified conjugate gradient method has been applied, Viallix and co-workers \cite{Viallix1988} computed Bloch walls and Bloch lines in ferromagnetic films. Alouges and co-workers \cite{Alouges2003} computed equilibrium configurations and switching fields of small ferromagnetic particles. 

Recently, steepest descent methods have been revisited for large scale minimization problems.\cite{Fletcher2012} 
In combination with a special choice of step-length \cite{Fletcher2005} steepest descent methods might even out-perform preconditioned conjugate gradient methods. 

In this work, we use a variant of these newly developed steepest descent methods and apply it to finite element micromagnetics. In order to take into account the constraint on the magnitude of the magnetization we use a curvilinear search\cite{Goldfarb2009} approach on the sphere. We compare the performance of the minimization algorithm with fast LLG solvers for computation of the hysteresis loop of permanent magnets. Further, we report on the graphic processor unit (GPU) implementation of the algorithm. 

\section{Method}
We are interested in finding local minima of the total Gibb's free energy, i.e. we want to solve
\begin{align}\label{min_en}
\min \phi_{t}(\boldsymbol{m}) \quad \text{s.t.}\quad \left\|\boldsymbol{m}\right\| = 1, 
\end{align} 
where the total energy $\phi_t$ consists of stray field, exchange, anisotropy and external energy.\\
\subsection{Steepest descent direction}
Let $g:\,\boldsymbol{m} \mapsto \boldsymbol{m}/\left\|\boldsymbol{m}\right\|$ denote the map onto the unit sphere. We get for the variational derivative of $h:= \phi_t \circ g$ for unit magnetization, i.e. $\left\|\boldsymbol{m}\right\| = 1$, 
\begin{align}\label{steepest_acent}
\frac{\delta}{\delta \boldsymbol{m}} h(\boldsymbol{m}) = J_{g}^T(\boldsymbol{m})\frac{\delta}{\delta \boldsymbol{m}} \phi_t\big(g(\boldsymbol{m})\big) = \\ \frac{\delta}{\delta \boldsymbol{m}} \phi_t(\boldsymbol{m}) -  \big(\boldsymbol{m} \cdot \frac{\delta}{\delta \boldsymbol{m}} \phi_t(\boldsymbol{m})\big) \notag \boldsymbol{m},
\end{align}
where $J_g(\boldsymbol{m})$ is the Jacobian of $g$ at $\boldsymbol{m}$.\\ 

The steepest ascent direction $\frac{\delta}{\delta \boldsymbol{m}} h(\boldsymbol{m})$
corresponds to the orthogonal projection of $\frac{\delta}{\delta \boldsymbol{m}} \phi_t(\boldsymbol{m})$ onto the orthogonal complement of $\boldsymbol{m}$. 
Using \textit{Lagrange's formula}, i.e. 
\begin{align}
 \boldsymbol{a} \times (\boldsymbol{b} \times \boldsymbol{c}) = (\boldsymbol{a}\cdot \boldsymbol{c})\boldsymbol{b} - (\boldsymbol{a}\cdot \boldsymbol{b})\boldsymbol{c},
\end{align}
we can rewrite Eqn.~\eqref{steepest_acent} 
\begin{align}
\frac{\delta}{\delta \boldsymbol{m}} h(\boldsymbol{m}) =  \boldsymbol{m} \times (- \boldsymbol{m} \times \frac{\delta}{\delta \boldsymbol{m}} \phi_t(\boldsymbol{m})).
\end{align}

Assume now that $\phi_t$ is a spacial approximation of the total Gibb's free energy, e.g. by finite element discretization. By the definition
\begin{align}
H(\boldsymbol{m}) := - \boldsymbol{m} \times \nabla \phi_t(\boldsymbol{m}),
\end{align} 
a steepest descent method would calculate a new iteration $\boldsymbol{m}^{n+1}$ from a given normalized approximation $\boldsymbol{m}^n$ by
\begin{align}
\boldsymbol{m}^{n+1} = \boldsymbol{m}^{n} - \tau_n \, \boldsymbol{m}^n \times H(\boldsymbol{m}^n),
\end{align}
for a certain step size $\tau_n$. Note that the new approximation $\boldsymbol{m}^{n+1}$ is not normalized in general.
\subsection{Curvilinear search on the sphere}
In our iteration method we replace the steepest descent direction by
\begin{align}
- \frac{\boldsymbol{m}^n + \boldsymbol{m}^{n+1}}{2} \times  H(\boldsymbol{m}^n),
\end{align}
yielding the iteration scheme 
\begin{align}\label{curvilin}
\boldsymbol{m}^{n+1} = \boldsymbol{m}^n - \tau \frac{\boldsymbol{m}^n + \boldsymbol{m}^{n+1}}{2} \times  H(\boldsymbol{m}^n).
\end{align}
This update scheme preserves the length of the iterates, i.e. $\left\|\boldsymbol{m}^{n+1}\right\| = \left\|\boldsymbol{m}^n\right\|$, which can be easily checked by multiplying Eqn.~\eqref{curvilin} by $\boldsymbol{m}^n + \boldsymbol{m}^{n+1}$.
Moreover, the new state $\boldsymbol{m}^{n+1}$ can be computed by explicite formulas \cite{Goldfarb2009}.\\
We want to stress, that the update scheme \eqref{curvilin} can be seen as an implicite integration rule for the flow equation
\begin{align}
 \partial_t \boldsymbol{m} = -\boldsymbol{m} \times ( -\boldsymbol{m} \times \frac{\delta}{\delta \boldsymbol{m}} \phi_t(\boldsymbol{m})),
\end{align}
which is, up to constants, the LLG equation with only damping, since the effective field is defined as $\boldsymbol{h}_{\text{eff}} = - \delta \phi_t/\delta\boldsymbol{m}$.
\subsection{Step length selection}
The first step size $\tau_0$ is calculated by an inexact line search and all subsequent steps $\tau_n$ by the so-called \textit{Barzilian-Borwein} (BB) rule \cite{barzilai1988}. We therefore define $\boldsymbol{g}^n := \nabla h(\boldsymbol{m}^n) = \boldsymbol{m}^n \times (- \boldsymbol{m}^n \times \nabla \phi_t(\boldsymbol{m}^n))$, $\boldsymbol{s}^{n-1}:= \boldsymbol{m}^n -  \boldsymbol{m}^{n-1}$ and 
$\boldsymbol{y}^{n-1}:= \boldsymbol{g}^n -  \boldsymbol{g}^{n-1}$. The step size $\tau_n$ is determined such that $D^n := \tau_n^{-1} I$ is an approximation of the Hessian of $h$ at $\boldsymbol{m}^n$, i.e. the \textit{secant equation} $D^n \boldsymbol{s}^{n-1} = \boldsymbol{y}^{n-1}$ holds. The two possible solutions to this equation are
\begin{align}
\tau_n^1 = \frac{(\boldsymbol{s}^{n-1})^T \boldsymbol{s}^{n-1}}{(\boldsymbol{s}^{n-1})^T \boldsymbol{y}^{n-1}}, \quad \tau_n^2 = \frac{(\boldsymbol{s}^{n-1})^T \boldsymbol{y}^{n-1}}{(\boldsymbol{y}^{n-1})^T \boldsymbol{y}^{n-1}}.
\end{align}
One possibility is to alternately switch between $\tau_n^1$ and $\tau_n^2$. However, in our numerical tests we used the more elaborate strategy proposed for step selection in gradient projection methods \cite{loris2009}.\\
Note that the BB rule yields a non-monotonic method; as globalization strategy, we therefore used inexact line search if the new computed energy was still larger than the previous $20$ energies.    
\section{Results}
\subsection{Comparison with LLG solver}
Here we compare the steepest descent solver with a state-of-the-art finite element solver. We computed the demagnetization curve of a spherical Nd$_2$Fe$_{14}$B
particle with a diameter of 20 nm. The number of tetrahedral elements is 80,000. The edge length of a tetrahedron is 0.76 nm. Owing the small particle size and the fine mesh the problem is stiff. Stiffness in micromagnetics has been identified by the collaborative motion of many magnetic moments \cite{Torre1993} at once.
Generally, stiffness leads to slow convergence when explicit solvers are used. 

The steepest descent solver and the LLG solver compute the magnetostatic field via a magnetic scalar potential. The LLG solver uses a hybrid finite element / boundary element method for the computation of the magnetostatic field. The boundary element method is accelerated using hierarchical matrices. For ease of porting the software to the GPU, we use a space transformation method \cite{Brunotte1992} instead of the accelerated boundary element method for the treatment of the open boundary problem in the steepest descent solver. In the LLG solver we set the torque term to zero. Time integration is performed using a preconditioned, implicit method with adaptive time step selection \cite{Suess2002}.

The external field is ramped from 0 to -6T in steps of 0.01T. In the static solver the external field is decreased when  $\nabla h^T \nabla h<10^{-12}$, cf. Eqn.~\eqref{steepest_acent}. In the dynamic solver the field is changed at a rate of $-0.01$~T/ns.  
The field is applied at 45 degrees with respect to the uniaxial anisotropy axis. 

\begin{table}
\centering
\begin{tabular}{l c c c}\hline
method & error   & evaluations & CPU time (s) \\ \hline \hline
steepest descent & $6\times10^{-4}$ & 13,250 & 3,801 \\
LLG solver       & $2\times10^{-3}$ &  6,379 & 8,114 \\ \hline \hline
\end{tabular}
\caption{\label{tab:time} Comparison of the steepest descent solver and the LLG solver: Relative error in the coercive field, number of function evaluations, time to solution on a netbook with an AMD E2-1800 processor (1.7 GHz). The CPU time does not include the setup time (matrix assembly and LU decomposition).}
\end{table}

Table~\ref{tab:time} compares the time to solution and the number of  function evaluation for two algorithms. The function evaluations denote the energy gradient evaluations for the steepest descent method and the effective field evaluation in the LLG solver.  The computed coercive field is close to the analytic value (half the anisotropy field) for both simulation methods. Although the LLG solver requires less function evaluations, the steepest descent method is faster by a factor of $2.13$. In the steepest descent method, more then 90 percent of the time is spent for solving the linear system for the magnetic scalar potential.  

\begin{table}
\centering
\begin{tabular}{l |  l }\hline
computation of & linear algebra operation \\ \hline \hline
energy gradient  & sparse matrix  vector multiplication \\
search direction & vector add and multiply\\
step length & dot product \\
scalar potential & solve linear system \\ \hline \hline
\end{tabular}
\caption{\label{tab:tasks} Main tasks in the steepest descent solver.}
\end{table}

\subsection{GPU implementation}
The steepest descent method was also implemented on a graphics processor unit (GPU). The implementation of the steepest descent method is based on  sparse matrix operations and basic linear algebra.  The required linear algebra operations are listed in Table \ref{tab:tasks}. We use OpenBLAS for the single core CPU implementation  and the PARALUTION library \cite{Lukarski} (version 0.3.0) for  GPU implementation. 

We compared the time for computing the demagnetization curve. The CPU was an Intel i7, the GPU was a NVIDIA m2070. We obtained a speed-up of 4.8 on the GPU as compared to the fastest single core CPU implementation.  
\section{Conclusion}

The hysteresis loop of permanent magnets can be computed using energy minimization techniques. Steepest descent methods with modified Barzilai-Borwein step length selection outperform state-of-the-art LLG\ solvers. With fast sparse matrix libraries the proposed solver  shows reasonable performance on graphic cards.

\section*{Acknowledgments}
Work supported by the Austrian Science Fund (F4112-N13).

\end{document}